\documentclass[11pt]{article}
\usepackage[T1]{fontenc}
\usepackage[latin9]{inputenc}
\usepackage[a4paper]{geometry}
 \usepackage{verbatim}
\usepackage{float}
\usepackage{textcomp}
\usepackage{amsthm}
\usepackage{amsmath}
\usepackage{graphicx}
\usepackage{gensymb}
\usepackage[usenames,dvipsnames]{xcolor}
\usepackage{array}
\usepackage{color}
\usepackage{xr}
\newcommand{\duline}[1]{\underline{\underline{#1}}}

\newcommand{\hl}[1]{\textcolor{black}{#1}}

\begin{document}
\vspace{5mm}

{\LARGE Inferring coarse-grain histone-DNA interaction potentials

from high-resolution structures of the nucleosome} \vspace{5mm}

Sam Meyer$^{1,2,3}$\footnote{Contact: sam.meyer@ens-lyon.org} and Ralf Everaers$^1$\vspace{2mm}\\
\small{
$^1$ Universit\'e de Lyon, Laboratoire de Physique and Centre Blaise Pascal, Ecole normale sup\'erieure de Lyon, UMR CNRS 5672, Lyon, France\\
 $^2$ Universit\'e de Lyon, INSA-Lyon, INRIA, LIRIS, CNRS UMR 5205, Lyon, France\\
 $^3$ Universit\'e de Lyon, Microbiologie Adaptation et Pathog\'enie, INSA-Lyon, CNRS UMR 5240, Lyon, France}

\vspace{5mm}

This is an author-created, un-copyedited version of an article accepted for publication in Journal of Physics: Condensed matter. IOP Publishing Ltd is not responsible for any errors or omissions in this version of the manuscript or any version derived from it.

\begin{abstract}

The histone-DNA interaction in the nucleosome is a fundamental mechanism of genomic compaction and regulation, which remains largely unkown despite a growing structural knowledge of the complex. Here, we propose a framework for the extraction of a nanoscale histone-DNA force-field from a collection of high-resolution structures, which may be adapted to a larger class of protein-DNA complexes. We apply the procedure on a large crystallographic database extended by snapshots from molecular dynamics simulations. The comparison of the structural models first shows that, at the sites of histone-DNA contact, the DNA base-pairs are locally shifted outwards, consistent with locally repulsive forces exerted by the histones. In a second step, we show that the various force profiles of the analyzed structures derive locally from a unique, sequence-independent, quadratic repulsive force field, while the sequence preferences are entirely due to the internal DNA mechanics. We thus obtain the first knowledge-derived nanoscale potential for the histone-DNA interaction in the nucleosome. The conformations obtained by relaxation of nucleosomal DNA with high-affinity sequences in this potential accurately reproduce experimental values of binding preferences. We finally address the more generic binding mechanisms relevant to the $80\%$ genomic sequences incorporated in nucleosomes, by computing the conformation of nucleosomal DNA with sequence-averaged properties. This conformation is found to differ from those found in crystals, and the analysis suggests that repulsive histone forces are related to a local stretch tension in nucleosomal DNA, mostly between successive contact points. This tension could play a role in the stability of the complex.

\end{abstract}

%\pacs{8715kj,8715La,8716Sr}
%\submitto{\JPCM}
%\maketitle

\section*{Introduction}

The nucleosome is the basic unit of DNA compaction in the eukaryotic nucleus~\cite{Kornberg74}. Its central part, the Nucleosome Core Particle (NCP), incorporates around 147 base-pairs (bp) of negatively charged DNA, helically wrapped around an octamer of cationic histone proteins, forming an approximately two-fold symmetric complex~\cite{Luger97}. The physical mechanisms involved in this wrapping play an important role in the organization of the genome~\cite{Schiessel03} and the regulation of its expression~\cite{Flaus01}, and they have been studied for decades. 

A significant progress came from the crystallization of the NCP, which revealed its structural details at almost atomic resolution~\cite{Davey02}, and in particular the existence of 14 regularly spaced histone-DNA contact points, which were interpreted as the sites of non-specific interaction~\cite{Richmond03}. However, the structural information alone is insufficient to infer the location of strong interactions, let alone their physical nature. How do the mechanical forces distort the DNA at the contact points? Is there an electrostatic attraction in the intermediate regions where the DNA is more distant to the proteins? What is the elastic cost of wrapping the DNA in the core? \hl{Similarly, it is unclear, if the presently available structures reveal artifacts of strongly binding DNA sequences or universal features of DNA wrapping in nucleosome core particles. }

To answer these questions, it is useful to introduce physical models in the analysis of the structural data. Such models can be very different, depending on the characteristic length- and timescales of the considered problem. In a Molecular Mechanics model, all atomic details are taken into account, but the accessible sampling times (<~1~$\mu$s) are incompatible with nucleosome-scale events such as breathing ($\sim$ 10 ms)~\cite{Li05}. At the opposite extreme, DNA can be modeled as a uniform semi-flexible polymer, with the histone-DNA interaction taken typically proportional to the adsorbed length~\cite{Kulic03b,Biswas12}, but the very low resolution of this model makes it more suitable to the study of polynucleosomes~\cite{Wedemann02} than to the internal mechanics of the NCP. Here, we place ourselves at an intermediate, nanoscale, level of description, which we consider appropriate for the study of the NCP. The sequence-dependent structure and elasticity of DNA are described with the rigid base-pair model, which is of common use in mechanical models of the nucleosome~\cite{Anselmi00,Morozov09,Deniz11,Battistini12}. The difficulty is that while the internal DNA mechanics has been parametrized from a combination~\cite{Becker06} of experiment-based~\cite{Olson98} and simulation data~\cite{Lankas03}, there is no corresponding knowledge of the histone-DNA interaction involved in the wrapping \hl{at the considered lengthscale. The latter depend on subtle electrostatic effects, which are still difficult to simulate ~\cite{Mukherjee11,Ettig11} because of long equilibration times~\cite{Lavery14}, and were mostly investigated using lower resolution theoretical descriptions~\cite{Cherstvy05,Sereda10,Boroudjerdi13}.}

\hl{To circumvent this limitation, two very different approaches were proposed. Most studies simply considered that all sequences wrap the histone core in exactly the same conformation, which was then taken from crystallographic data or from smoother models~\cite{Anselmi00,Sereda10,Deniz11,Battistini12,DeSantis13}. This scheme was motivated by the non-specificity of nucleosome binding, which cover as much as~$\sim 80\%$ of the genome. However, the absence of specificity signifies that the histone-DNA \emph{interaction potential} (and not the conformation) is sequence-independent, consistent with the absence of direct contacts between the histones and the bases in the high-resolution structures~\cite{Richmond03,Olson11}. Even in this case, the DNA structure (and wrapping energy) is expected to depend on the sequence, due to the contribution of internal DNA mechanics. The extrapolation of any specific conformation to all sequences therefore has little theoretical support, as emphasized by the observation that the crystallographic conformations already differ significantly from one another~\cite{Olson11}, and likely even more with respect to those relevant in solution.} To overcome the limitations of this ``rigid'' approach, other authors~\cite{Morozov09} constructed an idealized model of the interaction, as a homogeneous finite quadratic spring directed toward the regular superhelical path of the DNA. This scheme crucially improves the description of the mechanics by allowing the DNA to adopt a sequence-dependent conformation; however, the experimental structures present substantial deviations from the ideal helix~\cite{Tolstorukov07}, and the forces exerted by the histones are unlikely to follow an ideal homogeneous pattern. A more accurate description of the interaction should thus incorporate a level of experimental knowledge contained in the crystallographic data. This kind of approach has been proposed in a recent model~\cite{Fathizadeh13}, where the authors impose elastic springs on the histone-contacted phosphate groups of the DNA, which are then parametrized from the analysis of a crystallographic structural model. The potential thus still relies on the \emph{ad hoc} choice of a particular input model and a given combination of interaction sites. Here, we adopt a different standpoint, and propose to \emph{extract} a non-specific nanoscale force field by analyzing the whole available collection of high-resolution structural models, without any assumption on the location and strength of the forces. This force field then allows to estimate the properties of nucleosomes incorporating DNA of arbitrary sequences. 

This kind of structural analysis is not \hl{new: Olson et \emph{al}. followed the same rationale to extract the internal elastic energy function of (naked) DNA at the nanoscale~\cite{Olson98}. Here we propose a self-consistent extension based on the analysis of the forces and torques experienced by the double-helix in NCP structures. As we have shown in Ref.~\cite{Becker09a} for the example of the rigid base-pair model, this information can be inferred from experimental input given a mechanical model of DNA. Here we repeat the exercise for the available high-resolution NCP crystal structures and record the forces and torques in different crystal structures as a function of the DNA position and orientation relative to the histone spool. The idea is then to reconstruct a sequence-{\em independent} histone-DNA interaction potential from the information on its derivative for a representative set of sequence-{\em specific} conformations. The resulting interaction potential is specific to (and optimally adjusted for use in combination with) the DNA mechanical model employed in its derivation. While we illustrate the approach for the rigid base-pair model, the self-consistent parameterization procedure histone-DNA potential can be applied to arbitrary coarse-grain elastic models of DNA~\cite{KnottsIV07,vSulc12,Gonzalez13}. }
%Here, we propose a self-consistent iteration, by extending the analysis to the external potential exerted by a complexed protein}. We extract DNA-histone interaction potentials from a collection of high-resolution NCP crystals by analyzing the forces and torques experienced by the double-helix as a function of its position and orientation relative to the histone. This mechanical information is not directly available from the experimental input, but can be inferred~\cite{Becker09a} from the DNA structure under the assumption that the DNA elasticity is described by a given (rigid-base pair) model. The resulting interaction potential is thus optimally adjusted for use in combination with the DNA mechanical model employed in its derivation.

The paper is organized in three sections. In the Models and Methods section, we present the background and the theoretical framework for the analysis of an ensemble of NCP structural models, {in a formulation that may easily be adapted to other non-specific DNA-binding proteins}. We also present the collection of structural models used in the analysis, which includes not only most published crystallographic structures, but also a collection of snapshots from MD simulations of the entire nucleosome, which we systematically compare with the experiment-derived models. In the Results section, we compare the structures and inferred forces of these models, and we show that these forces derive from a unique histone force field and we use it to relax nucleosomal DNA. Technical details of the employed procedure can be found in the Supplementary Material. \hl{The Discussion is then devoted to the most notable features of the extracted force field, and in particular the presence of repulsive forces at the histone-DNA contact points. We analyze the conformation of sequence-averaged nucleosomal DNA, and show that the stability of the complex could involve the presence of (local) stretch tension in the DNA, analogous to a stretched rubber band holding together a group of pencils.} The paper closes with a short Conclusion.

%%%%%%%%%%%%%%%%%%%%%%%%%%%%%%%%%%%%

\section{Models and Methods}
\label{sec:MM}

\subsection{Background}
\label{sec:back}

\paragraph{The rigid base-pair model of DNA}
\label{sec:RBP}

We describe the mechanical properties of DNA with the rigid base-pair (RBP) model. Neglecting its internal fluctuations, a base-pair is described as a rigid body with 6 degrees of freedom, $\underline{\xi}=\{ \boldsymbol\omega , \mathbf{r}\}$, where $\mathbf{\boldsymbol\omega}$ is the orientation and $\mathbf{r}$ is the position of the bp in the reference frame. The fluctuations of a bp-\emph{step} are then described by the 6-vector $\underline{\xi}$ representing the \emph{relative} orientation and position of the successive bp in the local frame~(tilt, roll, twist, shift, rise, slide), as defined by the conventional axes~\cite{Dickerson89}. The fluctuations of the successive steps are considered as \emph{independent}: the free energy contribution of a given step depends only on its conformation $\underline{\xi}$ and sequence $s$. Finally, we treat this free energy in the linear elastic approximation, \emph{i.e.} the free energy function is harmonic~: 
\begin{equation}
  \label{eq:harmonic}
  F(s,\underline{\xi})=(\underline{\xi}-\underline{\xi}_0(s))^t \ \duline{K}(s) \  (\underline{\xi}-\underline{\xi}_0(s))
\end{equation}
where $s$ is the step sequence, $\underline{\xi}_0(s)$ is the \emph{equilibrium conformation}, and $\duline{K}(s)$ is the 6x6 symmetric \emph{stiffness matrix} associated to the step.

\paragraph{Extraction of parameters from DNA structures}
\label{sec:olson}

The RBP model involves 21+6 parameters for each step sequence ($\underline{\xi}_0(s)$, $\duline{K}(s)$), which makes a total of 270 (considering self-symmetric steps). These parameters have been obtained from MD simulations~\cite{Lankas03,Lavery10}, from the analysis of a database of high-resolution crystallographic DNA or DNA-protein structures~\cite{Olson98}, or from combinations of both~\cite{Becker06}. Structural databases hold a large number of conformations $\{\underline{\xi}_i\}_{i=1,N}$ for each step sequence, from which it is possible to compute an average conformation and a covariance matrix. It is then assumed that these conformations follow a canonical distribution at an effective temperature, \emph{i.e.} the external constraints due to the crystal packing and the bound proteins act as the random forces of a thermal bath. The parameters of equation~\ref{eq:harmonic} are then simply given by the average conformation and the inverse of the covariance matrix computed on the dataset.

\paragraph{Nanomechanical analysis of nucleosomal DNA}
\label{sec:nanomechs}

The RBP model describes the \emph{internal} DNA elasticity. We now consider the deformation of DNA under an \emph{external} potential. In our system of interest, the NCP, the DNA is tightly wrapped around a histone octamer~\cite{Richmond03}. The aim of the nanomechanical analysis~\cite{Becker09b} is to infer, from the observed deformed shape of the DNA, the \emph{forces} experienced by the molecule. Conceptually, this is as simple as estimating a person's weight from the state of deformation of a scale.

\begin{figure}[t]
  \centering
  \hspace{6mm}\includegraphics[width=6cm]{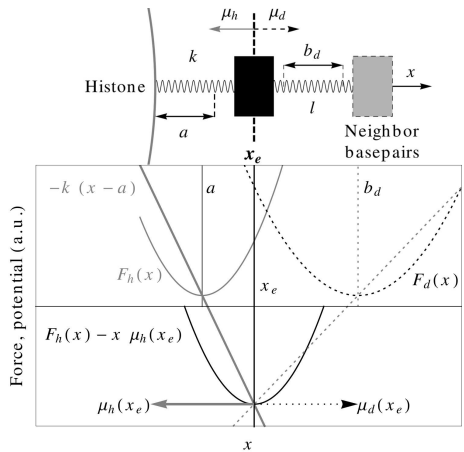}
  \caption{Equilibrium conformation of a nucleosomal DNA base-pair, in a unidimensional spring-like toy model of the nucleosome. \textbf{Upper panel:} Schematic depiction: the opposing forces $\mu_h$ exerted by the histone core (left, grey) and $\mu_d$ by the neighbour base-pairs (right, dashed) balance each other at the experimental conformation $x_e$. \textbf{Lower panel:} Corresponding schematic energy landscape in the simple case where the springs are linear (straight lines of mean force). Both free energy contributions are indicated by a parabola, and the equilibrium conformation $x_e$ is the minimum of the total free energy. }
  \label{im:wells}
\end{figure}

For simplicity, we illustrate the method on a unidimensional, spring-like toy model of the nucleosome on figure~\ref{im:wells}. 
The free energy function of the DNA RBP model (dashed parabola) is a potential of mean force~: in the regime of linear response, the \emph{mean} force required for a given \emph{mean} conformation is given by the derivative of the free energy~\cite{Becker09b} (dashed straight line). 
Under the hypothesis of mechanical equilibrium, the deformed conformation $x_e$ observed in the nucleosome is the result of the balance between this force $\mu_d$ and the external force $\mu_h$ exerted by the histone core (grey straight line). 
The knowledge of the DNA mechanical properties ($l$, $b_d$) and of the structure ($x_e$) implies that we can compute the force $\mu_h = -\mu_d$ responsible for the deformation. In practice, we first map the coordinates of the atomistic model into 6-dimensional RBP coordinates. The suitable derivative of the DNA free energy function at the observed deformed conformations $\underline{\xi}=\{\mathbf{\boldsymbol\omega} , \mathbf{r}\}$ of the successive steps then allows to compute the corresponding torques $t$ and forces $f$. 

Note that the method does \emph{not} require the potential to be harmonic: the condition of validity is that the regime of linear response is still valid in the deformed system. Whether this condition holds for the RBP in nucleosomal DNA remains an open question~\cite{Zhurkin13}: the base-pairs are strongly deformed, with possible changes of the backbone states and subtle electrostatic effects~\cite{Richmond03}. We note however that this model was successful in predicting the position of twist defects in the NCP~\cite{Becker09a} and is widely used to estimate the sequence-dependent nucleosome association free energies~\cite{Morozov09,Xu10}. Here, it is also justified \emph{a posteriori} by the relatively accurate predictions of sequence binding preferences (see Results).

%%%%%%%%%%%%%%%%%%%%%%%%%%%%%%%%%%%%%%%%%%%%%%%%%%%%%%%%%%%%%%%%%%%%%%%%%%%%%%%%
\subsection{Extraction of the coarse-grained force field in a biomolecular complex}
\label{sec:theory}

 \begin{figure}[t]
   \centering
   \includegraphics[width=8cm,trim=0 8mm 0 0]{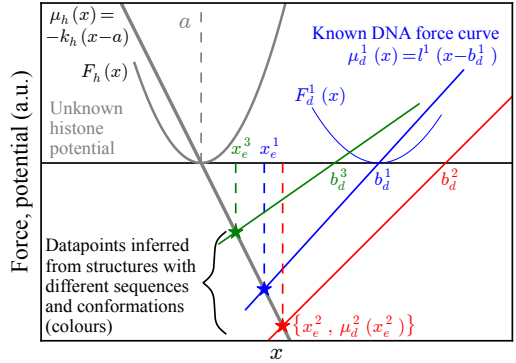}
   \caption{Determination of the parameters of the histone-DNA quadratic potential from a collection of structural models. The different observed conformations (stars) result from different DNA sequences/conformations (colours), corresponding to different lines of mean force (see figure~\ref{im:wells}). For each structural model, the nanomechanical analysis provides the coordinates $\{x_e,\mu_d\}$ of the star, and the 
elastic parameters $\{k_h,a\}$ of the histone potential (grey thick line) can then be estimated by linear regression. The same analysis can be carried on a more complex potential, provided the number of sampled datapoints is sufficient. }
   \label{im:fitpot}
 \end{figure}

The nanoscale elastic potential of internal DNA deformations was parametrized from the statistical analysis of an ensemble of high-resolution structural models~\cite{Olson98}. The employed scheme is only applicable if the available number of structural datapoints is much larger than the number of parameters in the potential: this constraint restricted its application to the RBP (rather than, \emph{e.g.} the rigid base) model of DNA. In the more complex NCP system, a simple estimation of the independent coordinates shows that the same method cannot be simply transposed. The total potential experienced by a given DNA bp is the sum of (i) the internal DNA elastic potential and (ii) the external potential exerted by the histones. They depend (i) on its position relative to its neighbours (2x6=12 degrees of freedom) and (ii) on its position relative to the histone core (6 d.o.f.)~: if the latter are fixed in the laboratory frame, these are the absolute coordinates. Even in the harmonic approximation, the fluctuations of the bp positions are therefore described by 18x18 stiffness matrices. Their parameters depend on the sequence of the considered bp and its neighbours~: altogether, a harmonic model thus involves 4752 parameters. As a comparison, there are currently $\sim 50$ published NCP crystal structures. 

We notice however that this considerable parameter set is largely redundant with the one of \emph{internal} DNA elasticity, already determined from independent experiments on naked DNA. In the nanomechanical analysis described in the previous paragraph, we showed that the combination of this prior knowledge with \emph{one} NCP high-resolution structure already allows to compute the \emph{static} forces acting on the DNA. Here, we go one step further, and show that the application of this analysis to a limited \emph{ensemble} of structural models yields the \emph{potential of mean force}. 

\begin{figure}[t]
  \centering
  \includegraphics[width=8cm,trim=0 0 0cm 0]{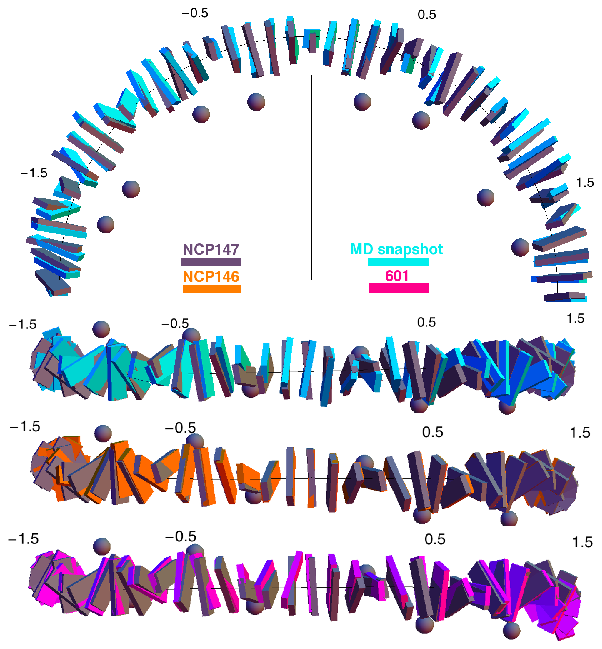}
  \caption{Comparison of the conformations in the crystallographic structures and a MD snapshot. Crystal structures: NCP147 (1kx5, grey), NCP146 (1kx3, orange) and 601 (3mvd, magenta). MD snapshot (cyan): MD1 (see table~\ref{tab:allstru}). The contact points are located at the semi-integral SHL, and the primary bound phosphates in the NCP147 structure are depicted as grey spheres.}
  \label{im:globstructure}
\end{figure}
\begin{figure}[t]
  \centering
  \includegraphics[width=9cm,trim=0 0 0 0mm]{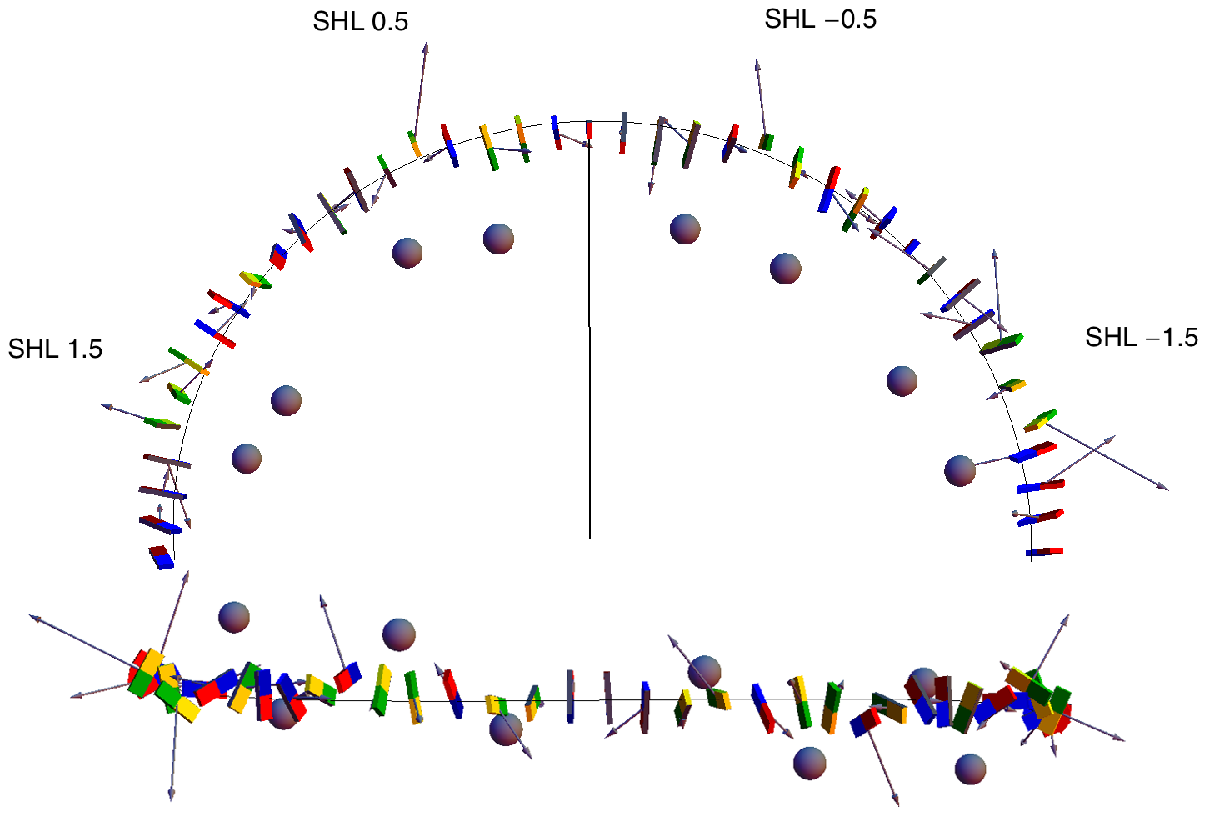}
  \caption{Force profile of the NCP147 structure (grey arrows): view along the superhelical axis (upper image) and along the dyad axis (lower image). At the contact points, the base-pairs are pushed outwards by strong forces, with a different pattern at SHL $\pm$0.5 and $\pm$1.5 (see text). Arrow size proportional to the force magnitude. Dyad axis indicated by a black line, and primary bound phosphates as grey spheres. Base-pair depiction scale 0.25. }
  \label{im:forstru_NCP147}
\end{figure}

The method is illustrated on figure~\ref{im:fitpot}, in the unidimensional toy model already used in the previous paragraph. The force inferred from each individual structure is the local derivative of the histone potential~: $\mu_h=-(dV_h/dx)|_{x_e}$. 
Importantly, while the DNA potential depends on the sequence and conformation of the analyzed structure, the histone potential $V_h$ is assumed to be non-specific. Each analyzed structure thus provides an independent point $\{x_e,(dV_h/dx)|_{x_e}\}$. Even a limited number of such datapoints allows to reconstitute the curve $(dV_h/dx)(x)$ and the potential of mean force from which it derives. Figure~\ref{im:fitpot} shows that, for a harmonic potential, the points align on a linear force curve and allow to fit the two parameters $\{a,k\}$. The same construction can be generalized to any functional form, provided the number of datapoints is sufficient. 

In the RBP model, each bp is described by a 6-dimensional vector $\underline{\xi}$ representing its position and orientation. The histone potential must therefore be fitted in the 6-dimensional configuration space. 
In the particular case of a harmonic potential,
\begin{equation}
  \label{eq:harmonicpot}
  V_h(\underline{\xi})=(\underline{\xi}-\underline{\xi}_0)^t \ \duline{K} \  (\underline{\xi}-\underline{\xi}_0) \quad, 
\end{equation}
the force $\mu_h$ depends linearly on the conformation $\underline{\xi}$: 
\begin{equation}
  \label{eq:fq}
  \mu_h(\underline{\xi}) = - \duline{K} \ (\underline{\xi}-\underline{\xi}_0)
\end{equation}
Here, the unknown histone potential is described by the 6-dimensional equilibrium conformation $\underline{\xi}_0$, and the 6x6 symmetric stiffness matrix $ \duline{K}$, \emph{i.e.} 27 parameters \hl{per base-pair}. These parameters are determined by fitting the datapoints $\{\underline{\xi}_i,\mu_i\}_{i=1,N}$ according to equation~\ref{eq:fq}, where each of the $N$ analyzed structural models contributes for 6 points. \hl{For the entire 79 bp long internal turn of the NCP on which we focus in the study, there are 27x79=2133 elastic parameters, and each structure contributes for 6x79=474 datapoints. Algebraically, 5 structures are sufficient to solve the minimization problem. However, considering that the 6 coordinates of each inferred force are not obtained independently, it is desirable to have at least 27 independent structures (here we have 118). }

%%%%%%%%%%%%%%%%%%%%%%%%%%%%%%%%%%%%%%%%%%%%%%%%%%%%%%%%%%%%%%%%%%%%%%

\subsection{Dataset of NCP structures}

\begin{figure*}[t]
  \centering
  %\begin{minipage}{.61\linewidth}
    \includegraphics[width=11.4cm,trim=7mm 0 40mm 0]{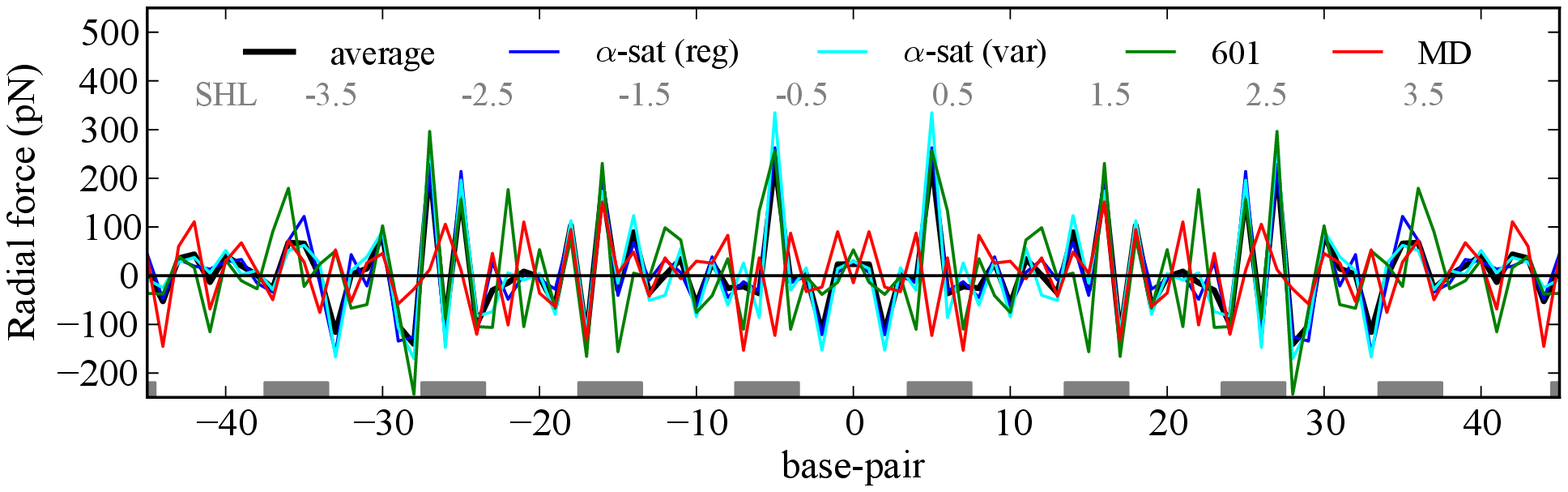}
  %\end{minipage}
  %\begin{minipage}{.38\linewidth}
    \caption{Radial force profile, averaged among the different groups of structural models.  A
      positive value indicates a force directed outwards: the contact
      points are therefore the locations of \emph{repulsive} forces,
      whereas attractive forces are weaker and less localized. Contact
      regions are indicated by grey rectangles.  }
    \label{im:rad}
  %\end{minipage}
\end{figure*}

\hl{The proper sampling of the interaction potential depends crucially on the available dataset, which should include a variety of conformational states, representative of different regions of the energy landscape. Here, our dataset is composed of three families of structural models (the detailed list is given in table~\ref{tab:allstru} in the Supp. Mat.). Most crystallographic structures (45) include derivates of the same human $\alpha$-satellite sequence~\cite{Luger97}, and 4 structures are based on the 601 sequence~\cite{Lowary98}. Throughout the article, we mostly refer to a few well-resolved structures, ``NCP147'' (PDB 1kx5) and ``NCP146'' (PDB 1kx3)~\cite{Davey02}, ``601'' (PDB 3mvd)~\cite{Makde10} and ``601L'' (PDB 3ut9)~\cite{Chua12}. A part of the dataset includes modified or variant histones, in which case the corresponding ``excited'' structure cannot be used to infer the ``regular'' nucleosome potential. However, this is true only in the region in contact with the perturbation, which is generally limited. We assumed that in the remaining part of the complex, the external potential is not modified. If the forces resulting from the perturbation partially propagate to remote locations through DNA mechanics, they might even allow to sample new (and higher) parts of the energy landscape of the nucleosome. Finally, the dataset includes snapshots from Molecular Dynamics runs of the entire nucleosome (excluding the histone tails) with the NCP147 sequence: 5 snapshots where the thermal fluctuations give access to other excited states, and 5 ``relaxed'' versions thereof, obtained through short energy minimization. The snapshots were separated by 2 ns in the MD trajectory. In the nanomechanical analysis of these snapshots, we implicitly assume that they represent local (metastable) mechanical equilibria (see figure~\ref{im:profiles} in Supp. Mat.). }

\hl{The histone octamer has an axis of symmetry, which passes through the central NCP bp (the dyad axis), and we therefore assumed that the histone potential should be symmetric with respect to this axis. The structural models are not symmetric however, either because the employed DNA sequence is non-palindromic (\emph{e.g.} 601), or because the two halves have crystallized in different conformations. We take advantage of this effect by considering the DNA conformation along either strand as independent data. This operation makes the dataset symmetric with respect to the dyad axis, while doubling the number of datapoints at each position (hence a total of 118 datapoints). The atomic coordinates of the different structures were mapped into RBP coordinates with the program 3DNA~\cite{Lu03}, and superposed on the reference NCP147 structure by minimizing the sum of distances between the base-pair centres. Note that we eliminated some additional crystallographic structures from the dataset, when the atomic coordinates could not be properly mapped into RBP coordinates, probably because of a higher level of noise. }

%%%%%%%%%%%%%%%%%%%%%%%%%%%%%%%%%%%%%%%%%%%%%%%%%%%%%%%%%%%%%%%%%%%%%%

\section{Results} 
\label{sec:results}

\begin{figure*}[t]
\includegraphics[width=17cm,trim=0 0mm 0cm 0]{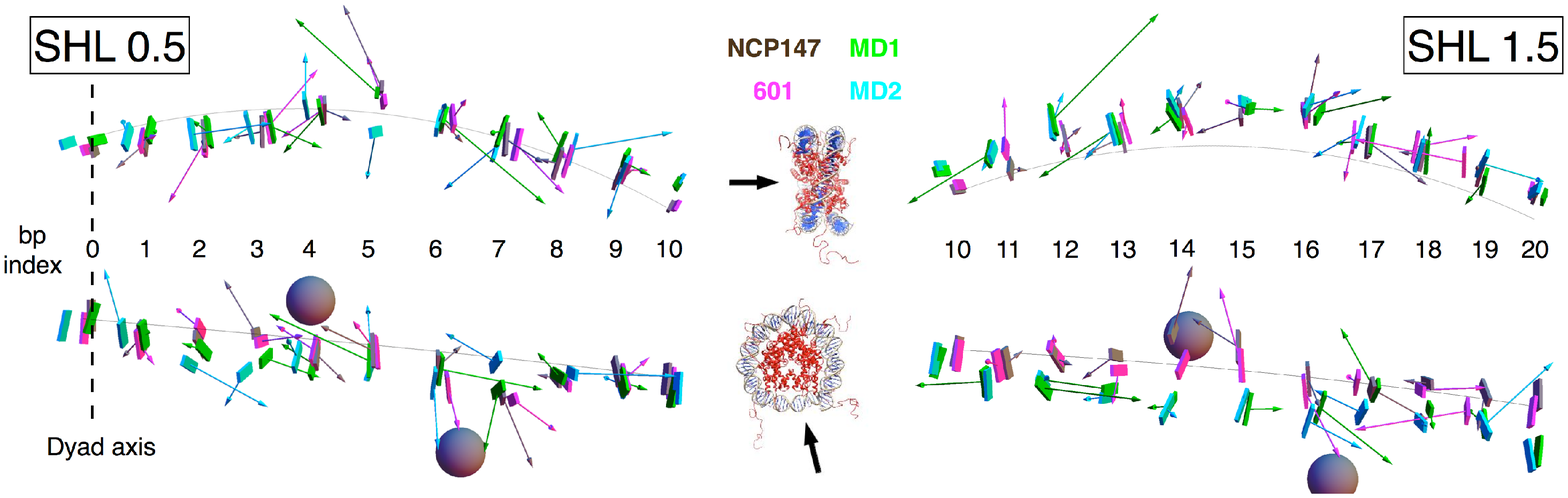} 
\caption{Base-pair conformations and inferred forces at the histone contact points SHL 0.5 (left) and 1.5 (right), in NCP147 (grey), the 601 structure 3mvd (magenta), and two MD snapshots: MD1 (cyan) and MD2 (green). View along the superhelical axis (upper panel) and in direction of the NCP (lower panel). The upper panel shows that the contacted bp are \emph{pushed away} from the core, consistent with a force directed \emph{outwards} (see details in text). }
   \label{im:forstru}
\end{figure*}

The nanomechanical analysis~\cite{Becker09b} of \emph{individual} high-resolution structural models of the Nucleosome Core Particle (NCP)~\cite{Davey02} showed that strong forces concentrate at the $\sim$10-bp periodic contact points with the histones, where stereotyped force motifs result in strong distortions of the DNA base-pairs~\cite{Richmond03}. 
In this section, we generalize this analysis to an \emph{ensemble} of high-resolution structural models of the NCP, following the method described in detail in the Models and Methods section. 

The analyzed dataset includes 49 structural models obtained from high-resolution X-ray crystallographs (see Models and Methods and the complete list in Supp. Mat., table~\ref{tab:allstru}). Most of these structures are based on the same $\alpha$-satellite sequence~\cite{Luger97} or related ones, including the ``NCP147'' structure~\cite{Davey02}, which is the best resolved nucleosomal structure and is used as a reference in the following. Because of this sequence similarity, the ability of this dataset to represent the whole conformational ensemble of nucleosomal DNA is questionable. Recently, 4 additional crystallographic structures \cite{Makde10,Vasudevan10,Chua12} were obtained with the strongly positioning 601 sequence~\cite{Lowary98}, whose structural features were found to differ from the previous ones~\cite{Olson11}: their inclusion in the dataset therefore increases the variety of the sample, and thus allows to sample new parts of the energy landscape of the nucleosome. 

In order to further increase this variety, we have included a set of structural models obtained from snapshots of all-atomic MD simulations of the entire nucleosome including the NCP147 sequence. Such simulations are becoming computationally tractable~\cite{Ettig11} and present different conformations than the knowledge-based models (see below). Whether these conformations are representative of actual nucleosomes remains an open question: possible bias include the limited sampling time (10 ns in our case) and the employed force fields. We therefore systematically compare the features observed in these models with those based on experimental data. \hl{The reader should however also keep in mind that the latter could be equally biased by the employed sequences, and by the crystallization process, with most structures crystallized in the same orthorombic geometry (albeit with different unit cell sizes). The relatively good agreement between the experimental and simulated models, and, \emph{e.g.}, the 601 structure that crystallized in a different geometry, suggests that these possible bias are not the dominant effects~\cite{Dickerson94}.}

\subsection{Base-pairs are shifted outwards at the contact points}

Figure~\ref{im:globstructure} shows four superhelical locations (SHL) ($\sim$ 1/2 turn) of the NCP, for the ``canonical'' NCP147 structure (PDB ID 1kx5, grey), which we will consider as a reference in the remaining analysis, and for (i) one of the MD snapshots (MD1, cyan), (ii) the NCP146 structure (PDB 1kx3, orange), which has the same sequence as NCP147 except at SHL -2.5 where it exhibits a twist defect, and (iii) a 601 crystal structure (PDB 3mvd, magenta). The contact points are located at semi-integral values of the SHL, where the minor groove faces the histone octamer: the primary bound phosphates~\cite{Richmond03} are here depicted by grey spheres. 

Overall, the bp in the MD snapshot remain remarkably close to those of the experimentally-derived structures. The deviations from the original structure are larger than in the NCP146 crystal, where they are hardly visible except in the extreme left part where the twist defect of this structure~\cite{Davey02,Becker09a} begins to appear. On the other hand, they are apparently not larger than in the 601 crystal~\cite{Olson11}, and present no obviously aberrant feature. If these discrepancies are indicative of actual alternative nucleosomal conformations rather than artefacts of the simulation, they may thus constitute a valuable source of structural information. 

For a more detailed analysis, we use a reduced depiction of the bp of the NCP147 structure on figure~\ref{im:forstru_NCP147}. Remarkably, the bp located at the contact points appear slightly \emph{shifted outwards} with respect to the average superhelical path (grey line), a counter-intuitive feature given that the contact points are generally considered as the points of \emph{attractive} interaction with the core histones. This feature is even more visible on the more detailed figure~\ref{im:forstru}, at bp 5 (SHL 0.5) and bp 14-16 (SHL 1.5), and confirmed by the distance profile with respect to the superhelical axis (figure~\ref{im:raddist} in Supp. Mat.).

\begin{figure*}[t]
  \centering
  \includegraphics[width=9cm,angle=90,clip=true,trim=2mm 2mm 2mm 2mm]{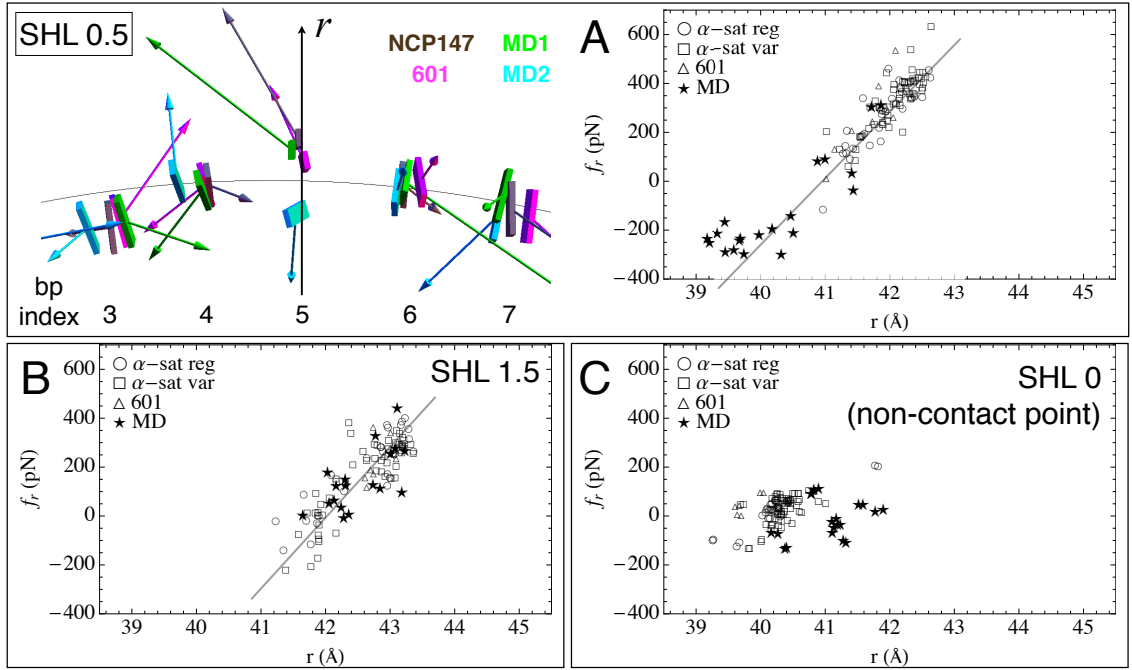}
  \caption{Relation between the bp radial position and the external force, at the SHL 0.5: example of the central bp +5.\textbf{ (A)} Same colours and conventions as on figure~\ref{im:forstru}. View along the superhelical axis, where the radial coordinate $x$ of (B) is vertical. \textbf{(B)} Force-displacement curve in the radial direction. Symbols: $\alpha$-satellite-based crystallographic structure with regular (circle) or variant (square) histones (see Mod. and Meth.), 601-based crystallographic structure (triangle), MD snapshot (star). The bp divide into two groups with either strong repulsive radial forces (most crystallographic structures, and some snapshots) or negative forces (most MD snapshots, and a single crystallographic structure). The same positive correlation is found for the whole dataset, corresponding to a repulsive quadratic potential centred at $x \simeq 3.75$ nm.}
  \label{im:corr}
\end{figure*}

\subsection{Strong and repulsive forces at the interaction sites}

While the crystallographic data give detailed structural information on the complex, the understanding of the underlying physical mechanisms requires the introduction of physical models. In the nanomechanical analysis~(see Models and Methods) we combine the rigid base-pair (RBP) model of DNA with high-resolution data, in order to infer the nanoscale forces acting on DNA in protein-DNA complexes~\cite{Becker09a,Becker09b}, under the hypothesis of mechanical equilibrium at the nanoscale. Here, we apply this procedure to the complete dataset. To regularize the force profiles and facilitate their comparisons, the structures were here allowed to slightly ``prerelax''~\cite{Becker09a} (see technical details in Supp. Mat.). 

The forces of the NCP147 structure are shown in 3D on figure~\ref{im:forstru_NCP147}. They are strongest at the contact points, where they exhibit characteristic patterns~\cite{Becker09a}: at SHL $\pm$0.5, a strong radial force acts on the central bp, and two approximately opposed forces 2 bp away on either side, resulting in a global torque on the chain. At SHL $\pm$1.5, two strong, approximately radial forces separated by 2 bp. Importantly, figure~\ref{im:forstru_NCP147} shows that the major forces point \emph{outwards}, consistent with the associated bp being pushed away from the core at the contact points. This is reflected in the radial component of the force profile, figure~\ref{im:rad}, which indicates that the contact points are the locations of \emph{repulsive forces}. These characteristics of the contact point forces are generic, and shared by all families of analyzed structures, even though their profiles differ in the details (see also figure~\ref{im:profiles} in Supp. Mat.). For a better comparison, it is useful to inspect these forces directly on the 3D structures, as depicted on figure~\ref{im:forstru}. 
In most cases, the contacted base-pairs (\emph{e.g.} 5 and 16) are indeed shifted outwards by repulsive forces (upper panel). In the transverse direction (lower panel), these base-pairs are also attracted in the direction of the primary bound phosphates. There are some exceptions however: as an example, in one of the MD snapshots (cyan) bp 5 is shifted inwards by a strong negative radial force. 

\hl{These qualitative observations already allow to separate the most generic features of nucleosome binding, which will be the object of the upcoming Discussion, from the specificities and irregularities inherent to each particular model. 
Anticipating the Discussion, we merely note that the unexpected repulsive forces are perfectly compatible with the stability of the complex, provided they are compensated by inwards internal forces resulting from a (local) stretch tension in the DNA. This is the situation of a common macroscopic example: that of a stretched rubber band holding together a group of pencils. In a next step, the large number of available structures gives the statistical power to quantify the differences between the analyzed structures}. Figure~\ref{im:corr} shows the radial component of the force, for all analyzed models, at the anchor points 0.5 (A) and 1.5 (B), and in an intermediate region (C). While the forces are weak and noisy in the latter case, they are stronger and well-defined in the former. At bp 5, while most models exhibit repulsive forces, we notice that the negative force observed previously in a single MD snapshot is in fact representative of a whole alternative group of datapoints (including a crystallographic structure). It may therefore constitute an alternate binding mode, rather than an artefact of the analysis.

\subsection{Sequence-dependent forces derive from a sequence-independent histone-DNA potential}

Figure~\ref{im:corr} shows that the forces, inferred from completely independent structures, are not randomly distributed: rather, they align on well-defined linear force-extension curves at the expected sites of histone-DNA interaction (A-B). The physical interpretation of this correlation is that these various forces derive from a unique histone-DNA potential of mean force: each inferred datapoint is therefore the derivative of this potential at the considered position (see the Models and Methods section). Importantly, since the analyzed structures incorporate different sequences, this observation signifies that the histone-DNA interaction in the nucleosome is non-specific. This property has been often hypothesized based on the absence of direct molecular contact with the bases, but indirect specificity (mediated by \emph{e.g.} subtle electrostatic effects) could never be ruled out. Here, the combined analysis of structural data gives strong support to the hypothesis. 

\hl{The \emph{slope} defined by the datapoints gives the stiffness of the local harmonic approximation of the histone potential (Equation \ref{eq:fq} in Models and Methods). In the next step, we compute this potential by fitting the datapoints. Importantly, while figure~\ref{im:corr} shows only one particular dimension (radial coordinate) of the datapoints, the non-radial features are known to play an important role in the mechanics of the nucleosome~\cite{Tolstorukov07}. Our fitting procedure therefore involves the full 6-dimensional conformations of the analyzed base-pairs. We compared the quality of the fit with an randomly generated ``control'' sample, which mimics a situation where the variation in basepair positions is due to pure noise~(Figure~\ref{im:quality}): the analysis confirms the presence of detectable histone forces in the inner turn of the nucleosome, while the latter are less well-defined in the outer parts of the complex. This feature may signify that the key interactions with the core histones are concentrated in the inner turn, consistent with the $\sim 80$-bp definition domain of high-affinity sequences~\cite{Lowary98}, while the external parts might be involved in more subtle and less generic interactions with the tails~\cite{Ettig11}. In the following, we will consider the potential only in the internal turn. The fit is also more precise in the contact point regions where the bp are more localized and the forces are stronger, than in the intermediate regions where the relative noise is important, as could be expected from figure~\ref{im:corr}. Still, even in these regions, many inferred forces contain a detectable non-random component (see all radial plots on figure~\ref{im:allrad} in Supp. Mat.), indicating that the histones also interact at distance with the DNA, probably through electrostatics~\cite{Sereda10}. }

\begin{figure}[t]
  \centering
  \includegraphics[width=8cm,trim=10mm 10mm 8mm 0]{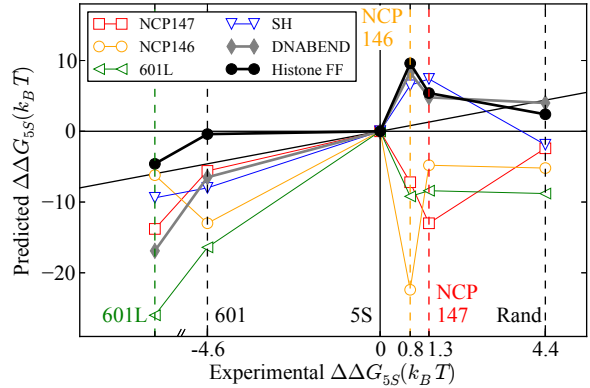}
  \caption{Model-derived vs. measured binding free energies differences of high-affinity sequences, for different mechanical models of the NCP. Rigid nucleosomal templates (coloured curves) fail to predict the sign of different sequences \hl{with respect to} 5S (right hand side). Our histone potential (black) reproduces qualitatively (correct ranking) but not quantitatively the experimental values, in line with a previous ideal model (DNABEND, grey). The experimental value of each sequence is indicated by a coloured tick or a label on the x-axis. } 
  \label{im:seqs}
\end{figure}

As a result of this step, we obtain the first nanoscale potential of the nucleosome based on experimental knowledge, and not on ideal models. As a consistency test, we computed the model-derived force profile along the 6 dimensions, as computed on the ``input structures'' NCP147 and the palindromic 601-derived structure 601L~\cite{Chua12}: their agreement with the original profiles validates the employed interpolation scheme (figure~\ref{im:forcecor} in Supp. Mat.). In the following section, we use this potential to estimate the sequence-dependent nucleosome wrapping elastic energy.

\subsection{Sequence-dependent conformation of nucleosomal DNA}

Our model describes the small harmonic fluctuations of nucleosomal DNA around its equilibrium position: it is therefore not directly applicable to large-scale rearrangements such as unwrapping or contact breaking. Rather, its main immediate application is the prediction of the sequence-dependent equilibrium conformation, which gives access to the elastic energy of wrapping the DNA around the core. Assuming that this elastic energy is the only sequence-dependent component of nucleosome binding (non-specific contacts, see previous section), the computed energies can be compared to the binding free energies, as measured in competitive binding experiments~\cite{Thastrom99,Thastrom04,Wu05,Morozov09}. 

Predicting these energies from mechanical models of the nucleosome has been already addressed in several studies, but in the absence of a reliable nanoscale histone-DNA potential, most of them simply threaded the DNA sequences onto fixed nucleosomal conformations, either the crystallographic models~\cite{Xu10} or ideal superhelical templates~\cite{Deniz11}. We addressed the question, how the predictions of such models depend on the chosen nucleosomal template. For the crystallographic models, a first difficulty arises from the presence of experimental noise, which results in a global factor affecting the computed energies and forces, as already noticed previously~\cite{Becker09a,Xu10a}: they are considerably larger than the experimental values (table~\ref{tab:seqs} in Supp. Mat.). We therefore estimated this global factor and, in the following, re-scaled all energies and forces accordingly (by 0.2, see Supp. Mat.).

The coloured curves of figure~\ref{im:seqs} compare the resulting predictions using different nucleosomal templates (NCP147, NCP146, 601L, ideal superhelix), for a small number of high-affinity sequences where the position of the NCP is well-defined. The variations between most rescaled datapoints are in the correct order of magnitude $\sim 10 k_B T$. However, and interestingly, each crystallographic structural template underestimates strongly \emph{the particular sequence} with which it was obtained, and poorly predicts the affinity of other sequences (as an example, all coloured datapoints on the right side of the plot have the wrong sign). Two conclusions can be drawn from these observation. Firstly, the employed DNA (rigid base-pair) elastic model is predictive of the sequences employed in the crystals, \emph{i.e.} it accurately describes the deformations present in these structures. This important observation justifies the choice of this model for the present study. Secondly, no \emph{single} structural template can properly describe the variety of conformations of even our limited set of sequences. This is true also for the ideal superhelix (blue), where some of these problems are absent, but which incorrectly predicts 5S to be less affine than a random sequence, suggesting that the binding mode of 5S differs from an ideal helix. Altogether, these observations support our efforts to \emph{combine} the different known structures into a single mechanical model. 

The thick black curve on figure~\ref{im:seqs} shows the predictions of our model, obtained by relaxing the different sequences in the histone potential computed in the previous section. In contrast to the previous computations, the relative preference of all sequences with respect to 5S are correctly predicted. This qualitative success may indicate a better description of the binding mode of 5S, as compared to all previous templates. Some properties of this unknown binding mode may be reflected in the structural profiles of the relaxed structures (Figs.~\ref{im:relstr} and \ref{im:relstr601} in Supp. Mat.), which indeed differ from all input structures. 

Interestingly, our results are also in relatively close agreement to the non-rigid mechanical model DNABEND~\cite{Morozov09}, where nucleosomal base-pairs are subject to forces directed toward the ideal superhelical path. Both models accurately rank the considered sequences, but fail to predict the quantitative values of the relative energies. However, we note that the validity of such a quantitative comparison is questionable: the measured free energies involve an ensemble of unwrapped and translated nucleosomal states, which are not considered in our model. These experiments have also been conducted on few sequences only, and the results depend strongly on the employed protocol~\cite{Thastrom99,Wu05}. To conclude, at the available qualitative level of comparison, our model successfully predicts the relative affinity of high-positioning sequences, and is consistent with the best existing mechanical models.

%%%%%%%%%%%%%%%%%%%%%%%%%%%%%%%%%%%%%%%%%%%%%%%%%%%%%%%%%%%%%%%%%%%%%%

\section{Discussion}

The mechanics of DNA in the nucleosomes has been previously described by two opposite approaches. Either a very specific conformation deduced from crystallographic data has been extrapolated to all genomic sequences~\cite{Anselmi00,Deniz11,Battistini12}, or the histone-DNA interaction was derived from ideal models which have limited experimental justification~\cite{Morozov09}. Here, we adopted an intermediate standpoint~\cite{Fathizadeh13}, and combined an extensive set of sequence-specific high-resolution NCP conformations with a mechanical model of DNA, to infer a sequence-independent nanoscale potential of the interaction. The proposal of this scheme is the central message of the present article. By construction, the extracted potential reflects the current knowledge, and results will certainly evolve with the refinement of the elastic model of DNA (for instance using a recently proposed rigid base model~\cite{Lavery10,Gonzalez13}) and particularly with the availability of additional NCP structures, which can then be incorporated into the input database. Such structural models may be obtained from experiments, but also from MD simulations of entire nucleosomes, which only begin to be computationally tractable~\cite{Ettig11}. 

\begin{figure*}[t]
  %\begin{minipage}{.6\linewidth}
    \centering
    \includegraphics[width=11.9cm,trim=3mm 0 0 0]{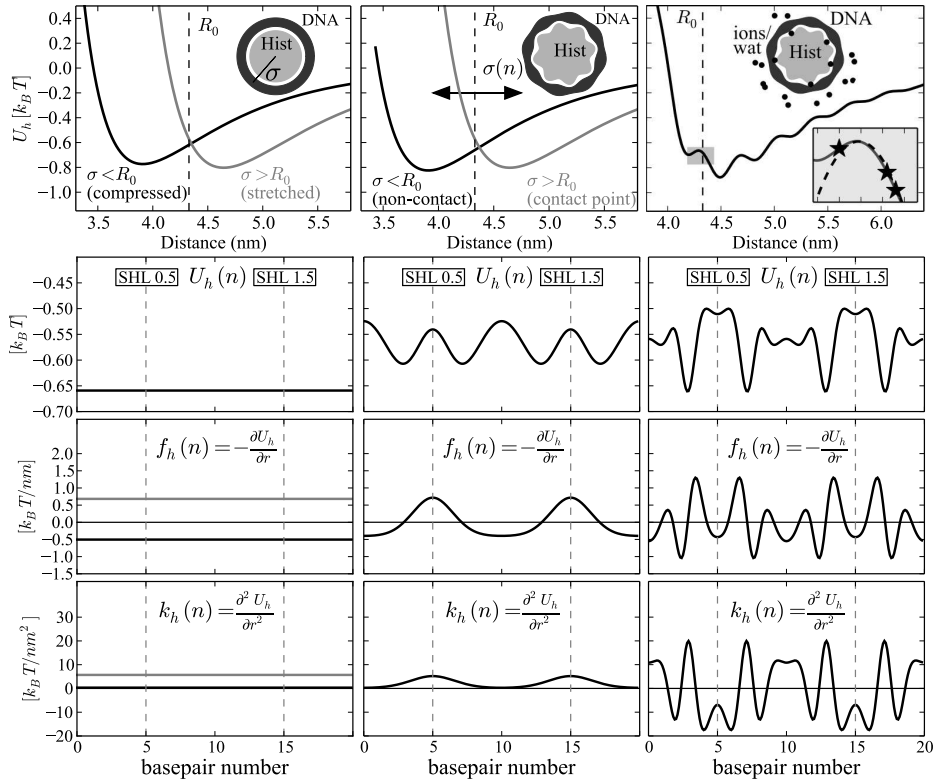}
  %\end{minipage}
  %\begin{minipage}{.4\linewidth}
    \caption{\hl{Toy elastic models of the nucleosome of increasing complexity (left to right, upper panel), with the profiles of the histone-DNA potential energy, radial force, and radial stiffness (upper to lower panels) along two helical turns of the DNA. These simplified models show that the stability of the complex is compatible with very different force and stiffness patterns, and in particular with repulsive forces, corresponding to a locally stretched DNA ($\sigma>r_0$). $\sigma$ is the effective DNA+histone radius, $r_0$ is the natural radius of wrapped DNA (defined by the average rise), and $n$ refers to the basepair index along the superhelical path (dyad at 0). A negative stiffness (lower right panel) can be expected to arise from the atom-scale irregularities at the contact points, in particular the layering of ions and water molecules. Functional form of the potential adapted from~\cite{Liang07}, with an arbitrary oscillation amplitude. }}
    \label{im:solvat}
  %\end{minipage}
\end{figure*}

While the detailed description of sequence-dependent effects may require further families of crystallographic structures, we now turn our attention to the most generic features exhibited by the analysis. The strongest forces involved in DNA wrapping are localized at the sites of histone-DNA contact, but, maybe surprisingly, these sites are not the location of radial attraction. Rather, in the analyzed structures, DNA base-pairs appear often shifted outwards by repulsive forces, which align on well-defined force-extension curves (figure~\ref{im:corr}). Just as surprising is the positive slope exhibited by these curves,  $f_x=+k_{xx} (x-x_0)$, which is the indication of a repulsive quadratic potential, $F_{xx}=-k_{xx} (x-x_0)^2/2$ (negative stiffness), in contrast to a regular spring where the slope would be negative (corresponding to a positive stiffness, see for instance figure~\ref{im:wells}). Note that this feature does not depend on the particular choice of coordinates; in the 6-dimensional configuration space where our analysis is conducted, a signature of this feature is that most fitted stiffness matrices have negative eigenvalues. %We therefore need to modify the schematic plots illustrating the model: Fig.~\ref{im:repwell} shows that the equilibrium points of several conformations align on a line of mean force compatible with our results. 
Importantly, this repulsive potential is compatible with the local mechanical stability of the base-pairs, provided it is compensated by the internal DNA elastic force (see figure~\ref{im:repforce1} in Supp. Mat.).

\begin{figure}[t]
  \centering
  \includegraphics[width=16cm,trim=0 0mm 0 0mm]{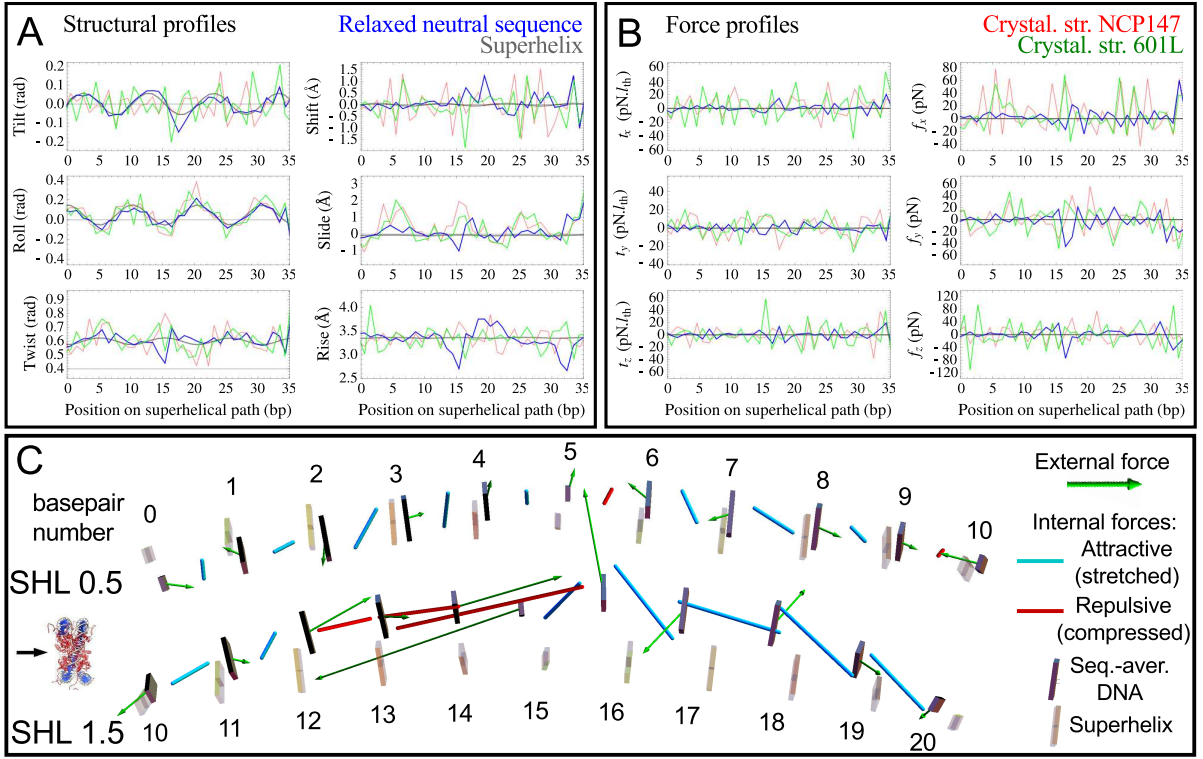}
  \caption{Structural and mechanical properties of sequence-averaged nucleosomal DNA. Profiles of (A) structural and (B) external force and torque components along the central turn of the nucleosome (blue), compared to the profiles of the superhelix (grey) and the crystallographic structures NCP147 (red) and 601L (green). The conventional axes are defined in the local frame, with $\hat{x}$ pointing toward the major groove and $\hat{z}$ along the helical axis~\cite{Dickerson89}. Structural forces rescaled by 0.2 (see text). (C) 3D depiction of relaxed sequence-averaged DNA, compared to the superhelix (semitransparent), along the superhelical axis. External forces in green, internal (tension) forces in cyan (stretched steps) and red (compressed steps). Each step exerts opposed internal forces on the two basepairs with the sign given by the depiction colour: $\mu= (\mathbf{t},\mathbf{f})=\pm \duline{K} (\underline{\xi}-\underline{\xi}_0)$ where $\underline{\xi}$ is the step 6-dimensional conformation, $\{\duline{K},\underline{xi}_0\}$ are the sequence-averaged stiffness and equilibrium step respectively (see Models and Methods). The generic nucleosomal DNA subject to histone forces deviates from the superhelix, but exhibits different features from the sequence-specific crystal conformations.}
  \label{im:neutralstr}
\end{figure}

\subsection{\hl{Repulsive forces are compatible with a stable complex involving locally stretched DNA}}

The existence of such repulsive potentials seems counter-intuitive, considering the expected electrostatic attraction between the positively charged histones and the negatively charged DNA backbone. We now suggest that these features result from the cylindrical geometry of the nucleosome, and may provide important information on the role of DNA \emph{stretch} in the physical mechanism of wrapping. To illustrate this point, we momentarily leave aside the employed complex multidimensional model of DNA where such effects are difficult to interpret, and we show that they already arise from the simplest ideal models of the nucleosome, shown on figure~\ref{im:solvat}. DNA is described as a uniform semiflexible polymer (worm-like chain) with bending and stretching rigidities, and the histone-DNA interaction is first described as a uniform Lennard-Jones-like potential accounting for the electrostatic attraction and short-range steric repulsion (left column). The wrapping mechanism then depends crucially on two lengthscales: (i) the interaction range $\sigma$, which corresponds to the added molecular radii of the histones and DNA, and (ii) the natural radius $r_0$ of the wrapped DNA length, imposed by the mean distance between successive base-pairs (rise). Importantly, if the actual radius differs from this spontaneous value, the resulting \emph{stretch}, rather than bending, dominates the elastic energy variations~(figure~\ref{im:stretchbend} in Supp. Mat.). \hl{Even in this oversimplified description, the grey curve of the left column ($\sigma>r_0$) illustrates a case where the mechanically stable solution (minimizing the total energy) is a stretched DNA conformation, which is held by \emph{repulsive} histone forces (positive value in the force panel). This scenario corresponds to the familiar example of a stretched rubber band holding together a group of pencils: at each contact point, the force exerted by the pencils is indeed repulsive, and compensated by the internal forces of the elastomer. }

\hl{While this simple analogy is useful to rationalize the observations, we now incorporate additional details in the models and show that they strongly affect the local shape of the interaction potential. In the middle panel of figure~\ref{im:solvat}, we consider the approximate 10-bp oscillatory structure of the nucleosome (sinusoidal $\sigma(n)$). For simplicity, the stiff DNA is assumed to keep a uniform radius, which minimizes the total energy integrated along the superhelical path. As a result, at the contact points, the protruding amino-acids hold the DNA by repulsive forces, while imposing a stretch tension to the molecule.} This tension also affects the intermediate parts, which experience a weaker attractive force, in qualitative agreement with many of our observations. The exception is the observation of a negative stiffness $k_h$ (positive slope of figure~\ref{im:corr}) at the contact points, \emph{i.e.} in the convex left region of the Lennard-Jones-like curve. This behaviour can be expected to arise from the atomic-scale features that are crucial in these regions. We suggest in particular that the molecular packing of ions and water molecules may play an important role, which is known to strongly influence macromolecular interaction in solution at short distances~\cite{Liang07}. This is especially relevant in these histone-DNA contact regions, where DNA is probably in direct competition with the first, strongly bound layer of counterions, and therefore subject to the so-called ``hydration'' or ``solvation'' forces~\cite{Marcelja11}. In this case, the effective interaction typically contains an oscillatory contribution, reflecting the ordered layering of ions and water molecules ($\sim 3$~\AA\ periodic), and decaying exponentially with distance~\cite{Liang07,Marcelja11}, as shown on the right column of figure~\ref{im:solvat}. The rugged potential leads to a stable complex, \emph{globally} very similar to the previous case, but exhibiting very irregular profiles of \emph{local} features, where all combinations of force and stiffness signs are possible, depending on the molecular details. Negative values of the stiffness correspond to local (concave) maxima of the potential, characteristic of the ionic or molecular shells. Interestingly, we already noted that in the datapoints of SHL 0.5~(figure~\ref{im:corr}A), not all structural models exhibit repulsive forces; in a second, well-separated group of inwards-shifted base-pairs, the force is attractive. We suggest that these groups could be located at the two sides of a peak in the solvent density function, corresponding to a positive and negative radial force respectively. They are indeed separated by a distance of $\sim 2$ \AA, remarkably close to the expected value if the attracted basepairs correspond to an alternate binding mode where the internal ion or water molecule is absent or displaced. This observation signifies that our coarse-grained DNA model might be able to detect atomic-scale features if many structures are analyzed simultaneously.

\subsection{\hl{Sequence-averaged nucleosomal DNA}}

\hl{In contrast to the usual view, the toy examples presented in the previous paragraph suggest that DNA stretching, in addition to bending, could play a key role in the mechanics of nucleosome wrapping. Such an effect can however not easily be detected directly on the crystallographic structural models, where the very specific employed sequences contribute, as well as the wrapping, to the observed deviations from a regular B-DNA conformation. Since we rather focus on the non-specific binding mechanisms, which are relevant to the $\sim 80\%$ genomic sequences that are incorporated in nucleosomes most of the time, we therefore propose to analyze in more detail the conformation of sequence-averaged (``neutral'') DNA, as obtained after relaxation in the histone potential, and where the uniform generic mechanical features allows to circumvent these problems. }
% show that the stability of the complex is compatible with many different radial force and stiffness patterns, in qualitative agreement with many of those observed in our data. This is in contrast with the usual view of nucleosome binding, where the forces between the histones and the wrapped DNA can only be attractive and compensate the tendency of the stiff molecule to unbend. In these examples, the repulsive forces are characteristic of a complex where the DNA is held in a locally stretched conformation, while the details depend on the rugged atomic-scale features at the contact points. Since such repulsive forces were observed at many different locations along the superhelical path, this effect could potentially play an important role in the wrapping. It 
%We checked the stability of the employed relaxation procedure by varying the initial state, 

%As a consistency test, the relaxation was carried with very different initial states, 
%We checked the robustness of the relaxation procedure by 
Figure~\ref{im:neutralstr} shows the profiles of the 6 step helical parameters (A) and force components (B) along the central turn of the nucleosome, for the sequence-averaged structure (blue line, the palindromic sequence makes it symmetric with respect to the dyad axis by construction). We see that this conformation is closer to the regular superhelix (grey) than to any family of crystallographic structures that we have analyzed (red and green), but still exhibits substantial deviations which are particularly apparent on the three-dimensional depiction of the structure (C). We checked that these deviations do not depend on the initial state employed in the relaxation procedure, but indeed reflect the properties of the extracted force field (figure~\ref{im:relaxcheck} in Supp. Mat.). Interestingly, while the local irregularities of the two sequence-specific families of crystallographic structures are quite different, those of the generic relaxed conformation generally deviate from both, confirming that their sequences may play an important role in these structural features. These non-superhelical features, maybe indicative of an alternate (and more generic) binding mode of the nucleosome, stem from the forces exerted by the histones, which are generally lower in the relaxed structure than in the input models, except at the contact points (particularly SHL 1.5). In the latter case where the steps are strongly deformed, some of the structural features may however also reflect the limitations of the employed quadratic models (either for the internal DNA mechanics or for the histone-DNA interaction)~\cite{Zhurkin13}: for instance, the strong rise peak at bp~16, which is absent of the original structures, might be due to the quadratic coupling with tilt, where a peak is indeed present. 

\begin{figure}[t]
  \centering
  \includegraphics[width=8.5cm,trim=0 3mm 0 0mm]{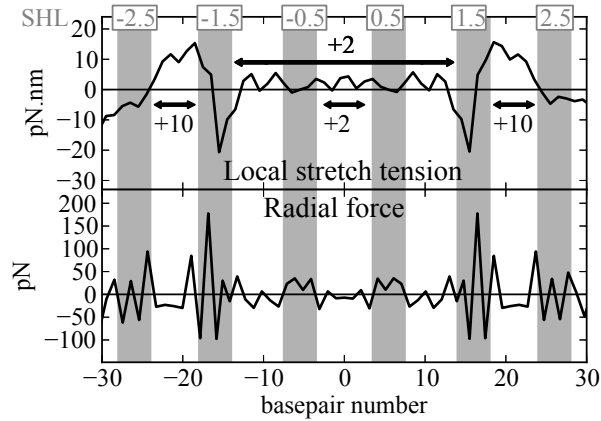}
  \caption{Profile of stretch tension (upper panel) and radial force (lower panel) along the central turn of sequence-averaged nucleosomal DNA. The contacted regions have a grey background. They are the location of strong, mostly repulsive forces, which result in a state of stretch tension in the intermediate regions where the radial forces are weak and mostly attractive. The average stretch energy per basepair is low in the central part ($\sim 2$ pN.nm), and much higher between SHL 1.5 and 2.5.  }
  \label{im:neutral}
\end{figure}

To investigate the possible role of internal mechanical tension in the energetics of the complex, figure~\ref{im:neutralstr}C shows not only the structure and the external forces (green arrows) exerted on the basepairs, but also the \emph{internal} forces which result from the deformations of the basepair steps. These forces are depicted as coloured bars, with the sign of the forces indicated by the colour, a red (resp. cyan) bar corresponding to a compressed (resp. stretched) step. In each case, two opposite forces act on the basepairs of the step, and tend to separate (resp. attract) them. These forces have very different directions and intensities along the path owing to the local deformations of the steps, but most of them correspond indeed to stretched steps. More precisely, there are strong irregular forces at the contact points (especially at SHL 1.5 where the steps are compressed) resulting from strong external (mostly repulsive) forces. In contrast, outside these regions, nearly all steps are stretched, while the external forces are weak. The internal forces have a strong radial component only in the vicinity of the contact points, while they become mostly tangential (\emph{i.e.} pure stretch) in the intermediate parts. The qualitative picture emerging from these observations is in agreement with the suggestions of the toy nucleosome models: the histones impose not only strong local (and irregular) deformations to the contacted steps, but also distance and orientation constraints between the successive contact points, which result in a state of tension of the helix in the intermediate regions where the external forces are weak. To quantify these observations, figure~\ref{im:neutral} shows the stretch component of the internal tension together with the radial external forces. Interestingly, the level of tension is very different for the different locations. The low repulsive forces at the contact points $\pm0.5$ correspond to a low intermediate average tension of $\sim 2$ pN.nm. But this tension possibly results from the propagation of the much stronger forces at SHL $\pm$1.5, which would distribute into the whole central region between bp -15 and bp +15 (average level 2 pN.nm). On the other side conversely, the tension concentrates between SHL $\pm 1.5$ and SHL $\pm 2.5$, with a strong level of $\sim$10 pN.nm. 

These observations demonstrate that DNA tension could play an important role in the mechanics of wrapping. Actually, this topic has been the object of a longstanding controversy~\cite{Ulanovsky83}, but the presence of a global stretch tension was never demonstrated, which may be partly due to experimental difficulties~\cite{Trifonov11}. In particular, we note that the tension was only assayed at the scale of the entire nucleosome, and for specific sequences which can substantially influence the properties of the DNA. Our results suggest however that the tension could affect only or mostly some portions of the complex, with a relevant extension scale of 10 bp corresponding to the distance between the successive contact points, and might thus be detected only at this resolution. Interestingly, if the net histone-DNA force is indeed repulsive at least at some contact points, then the DNA also ``pushes'' and holds the histones together by reaction, a feature qualitatively compatible with the observation that the octamer dissociates in absence of wrapped DNA~\cite{Schiessel03}. With our framework only describing the internal part of the complex, further investigation will be necessary to resolve these features along the entire superhelical path, but these observations already underline the strength of combining nanoscale mechanical models with structural data. They suggest to re-evaluate the role of DNA elasticity in the physics of the nucleosomal complex: instead of merely opposing the wrapping, it could contribute in its stability.

\section*{Conclusion}

The physical mechanisms of histone-DNA interaction in the nucleosome are a key ingredient of the genomic compaction in the nucleus, and yet they remain largely unknown. In this work, we have proposed a new method for the extraction of effective nanoscale potentials in DNA-protein complexes from the analysis of high-resolution structural data, and we have applied it to the NCP. We analyzed 49 crystallographic structural models, which are based on two families of high-positioning sequences. Because both types of conformations may differ significantly from the dynamic structures present in solution, this database was increased by the introduction of 10 snapshots from MD simulations of an entire nucleosome: we verified that their analyzed properties were generally consistent with those of knowledge-based models. In some cases however, they present differences, which may be indicative of alternate binding modes and were considered in the analysis. 

Within the rigid base-pair description of DNA, we find that the base-pairs are locally pushed away rather than attracted at the 10-bp periodic histone contact points. This behaviour can be described for the whole analyzed dataset by a repulsive quadratic force field at these locations. To test the validity of this knowledge-derived nanoscale potential, we compared the computed wrapping energy with measured binding free energies for a few high-positioning sequences, and found qualitative agreement, comparable to the best available estimates based on ideal models. The conformations obtained after relaxation differ often from both input families, and might indicate alternate conformations. %Our framework also allows to easily augment the analysis if new conformations are available in the future, or by the use of a more precise DNA physical model.

The extracted repulsive potential diverges from the most usual view of nucleosomal stability, where the electrostatic attraction opposes the bending stiffness of the stiff molecule. In the Discussion, we suggest to reconsider the role of DNA elasticity, which may \emph{contribute} in this stability instead of simply opposing the wrapping, if the DNA is constrained not only in the bending but also in the coupled twist/stretch degrees of freedom.

\section*{Acknowledgements}

We thank Richard Lavery for fruitful discussions, and Agnes Noy and Modesto Orozco (IRB Barcelona) who kindly provided the nucleosome snapshots. This work was supported by the Agence Nationale de la Recherche grant ``FSCF'' [ANR-12-BSV5-0009-01]. 

%\section*{References}

\bibliographystyle{iopart-num}
\bibliography{shortdot,biblio}

\end{document}